\documentclass[a4paper,11pt]{article}
\usepackage{jinstpub} 
\usepackage{lineno}
\usepackage{graphicx}

\newcommand{\GeCCD}{Ge-CCD}

\newcommand{\dut}{GEDI7\_-1\_0}                       
\newcommand{\el}{e$^{-}$}                             
\newcommand{\elpps}{e$^{-}$\,pixel$^{-1}$\,s$^{-1}$}  
\newcommand{\Fe}{$^{55}$Fe}
\newcommand{\Am}{$^{241}$Am}
\newcommand{\Ba}{$^{133}$Ba}
\newcommand{\Ka}{K$\alpha$}
\newcommand{\Kb}{K$\beta$}

\title{\boldmath Laboratory Characterization of a Ge-CCD Prototype for Future Near-Infrared Astronomical Instrumentation}

\author[a]{Emi Tamura}
\author[a]{Juan Cruz Estrada Vigil}
\author[a]{Prashansa Mukim}
\author[a]{Giovanni Pinaroli}
\author[b]{Christopher Leitz}
\author[b]{Ilya Prigozhin}
\author[b]{Kevan Donlon}
\author[c]{Santiago Ezequiel Perez}
\author[d]{Sven C. Herrmann}

\affiliation[a]{Instrumentation Division,
Brookhaven National Laboratory,\\
Upton, NY 11973-5000, USA}
\affiliation[b]{Lincoln Laboratory,
Massachusetts Institute of Technology,\\
Lexington, MA 02421-6426, USA}
\affiliation[c]{University of Chicago,\\
IL, USA}
\affiliation[c]{Cornell University,\\
NY, USA}
\emailAdd{etamura@bnl.gov}

\abstract{
Germanium charge-coupled devices (Ge-CCDs) extend CCD imaging beyond the
silicon cutoff and are candidate detectors for the near-infrared channel of
future astronomical instruments, in particular the spectrographs of
next-generation wide-field surveys. We report the laboratory characterization
at Brookhaven National Laboratory (BNL) of a prototype fabricated at MIT
Lincoln Laboratory. We describe the cryogenic test system and readout, and
present measurements of the output waveform, dark current, readout noise, and
X-ray response. The dark current at 94.69\,K is
$26.12 \pm 0.08$~\elpps. Under $^{55}$Fe irradiation without on-chip binning,
the charge generated by an X-ray is confined within $2\times2$ pixels,
and the Mn~K$\alpha$ line has a full width at half maximum (FWHM) of
$8.5 \pm 0.1$\% for single-pixel events. Since the charge generated
by X-rays above about 30\,keV spreads over $3\times3$ or more pixels,
the CCD was operated with $2\times2$ on-chip binning at higher energies.
This yielded FWHMs of $8.4\pm0.4$\% at 30.9\,keV ($^{133}$Ba,
ASCA grades 0, 2, 3, 4, and 6) and $7.45\pm0.04$\% at 59.5\,keV ($^{241}$Am, all grades).
These results establish a baseline for the device and for the BNL test system.
Measurements of the temperature dependence, optical response, and
charge-transfer performance will connect this baseline to the requirements of
near-infrared astronomical instruments.

}


\arxivnumber{8237739} 

\begin{document}
\maketitle
\flushbottom

\section{Introduction}
\label{sec:intro}

Future wide-field spectroscopic surveys aim to extend measurements of cosmic
structure and galaxy evolution to higher redshift. Their reach depends in part
on spectral features that shift beyond the wavelength range of silicon
detectors. Concepts for next-generation facilities, including the Stage-5
Spectroscopic Experiment (Spec-S5), therefore motivate the development of
detectors for a near-infrared channel \cite{besuner2025,schlegel2022}.

The scientific motivation for near-infrared detectors extends beyond
spectroscopy. A future Gaia-like astrometry mission, often referred to as
GaiaNIR, has been proposed to map stars through dust-obscured regions of the
Galaxy, including the Galactic center, bulge, and spiral arms. To retain
Gaia's scanning measurement approach, such a mission would require near-infrared
detectors capable of time-delay integration (TDI), or a comparable method of
low-noise charge-domain accumulation. Germanium detectors have been identified
as one possible technology for this application because they could combine
near-infrared sensitivity with charge-coupled device (CCD)-style charge transfer.
Large-format arrays with sufficiently low noise and high charge-transfer
efficiency remain important development
goals~\cite{hobbs2016gaianir,hobbs2019scanningnir,hobbs2021allskiynir}.

Near-infrared imaging has also been considered in the context of a possible
future infrared camera for the Legacy Survey of Space and Time (LSST)
facility after its nominal survey. A study
by the LSST Project Science Team identified germanium CCDs (Ge-CCDs)
as one possible route to
extend the red wavelength coverage, with an upper wavelength goal near
$1.8\,\mu\mathrm{m}$. This was a conceptual post-survey option, conditional on
detector cost and system-level changes that include the optics, thermal
background management, and readout electronics
\cite{kahn2019lsstfuture}.

Germanium is a candidate material for such detectors. Its smaller bandgap
than silicon can extend photoresponse into the short-wave infrared (SWIR), while
its compatibility with CCD architectures offers potential advantages including
charge-domain transfer, a small number of output amplifiers, and large
monolithic formats. These properties are attractive for both scanning
astrometry and wide-field imaging, although the requirements differ: TDI places
particular demands on charge transfer, while survey imaging also depends on
array scale, cost, thermal background, and optical throughput. Mission and
instrument studies treat Ge-CCDs as one candidate among several detector
approaches; low-noise operation, charge-transfer efficiency, backside
illumination, and scaling to large arrays remain development challenges
\cite{hobbs2019scanningnir,hobbs2021allskiynir,kahn2019lsstfuture}. MIT Lincoln
Laboratory and collaborators have demonstrated germanium CCD structures,
prototype imaging arrays, and X-ray measurements
\cite{leitz2017,leitz2020,donlon2025geccd}.

This paper reports the characterization of one of these prototypes in a test
system assembled at Brookhaven National Laboratory (BNL).
Section~\ref{sec:setup} describes the device, the
cryogenic system, and the readout. Section~\ref{sec:methods} describes the
measurement methods, and Section~\ref{sec:results} presents the dark current,
conversion gain, readout noise, and X-ray response.
Section~\ref{sec:discussion} discusses these results and places them in the
context of SWIR detectors for future instruments. The measurements reported
here are electrical or X-ray measurements;
we will report optical response in the near infrared in future work.

\section{Detector and Experimental Setup}
\label{sec:setup}

\subsection{Ge-CCD prototype}

 The device under test, designated \dut, is a germanium CCD fabricated at MIT Lincoln
 Laboratory and packaged for testing at BNL. It is a full-frame device
 with an imaging area of $256\times256$ pixels with a pixel size of $8.1\,\mu\mathrm{m}\times8.1\,\mu\mathrm{m}$.
 It is a front-illuminated, three-phase, p-channel device fabricated on an
 n-type germanium substrate, in which the signal charge is carried by holes.
 Only one of the two outputs can be used at a time.
 All of the measurements reported in this paper were taken using the MOSFET output.
 Details of the device are given by Donlon et al.~\cite{donlon2025geccd}.

\begin{figure}[htbp]
\centering
\includegraphics[width=0.70\linewidth, clip, trim={2cm, 10cm, 2cm, 9cm}]{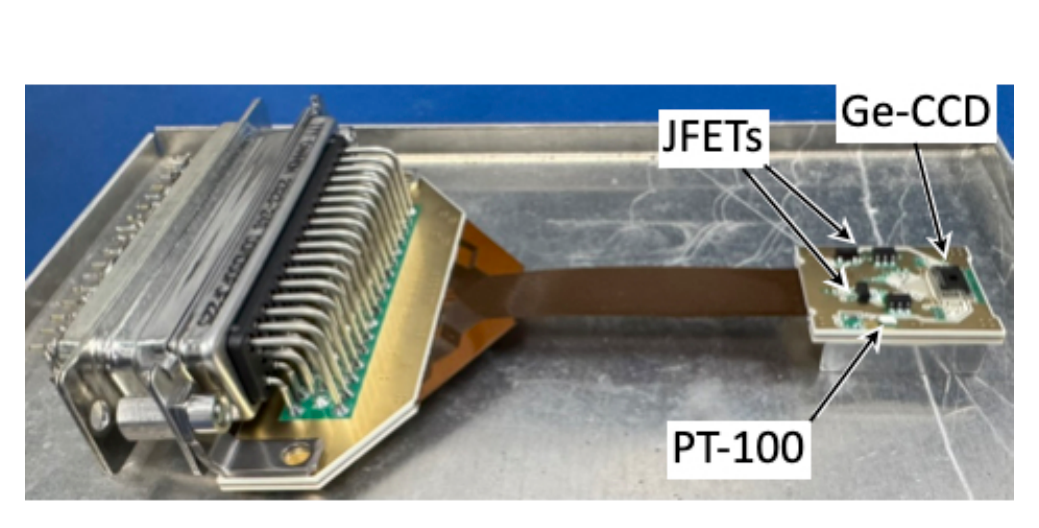}
\caption{\GeCCD\ prototype mounted on the interposer.
    The two outputs, one with an on-chip MOSFET amplifier and
    the other with a JFET amplifier, are each connected
    to a pair of JFETs on the interposer: a current-source
    load and a source-follower buffer.
    A 100\,$\Omega$ platinum resistance thermometer (PT-100)
    is mounted on the interposer to monitor the temperature
    near the CCD.}
\label{fig:packaged_geccd}
\end{figure}

 Figure~\ref{fig:packaged_geccd} shows \dut\ mounted on the custom interposer,
 to which it is die-bonded and wire-bonded. As in the MIT Lincoln Laboratory
 package~\cite{donlon2025geccd}, two JFETs are connected to each output. The temperature
 is monitored with a platinum resistance thermometer mounted on the interposer.

\subsection{Cryogenic test system}

\begin{figure}[htbp]
\centering
\includegraphics[width=0.70\linewidth, clip, trim={7cm, 5cm, 7cm, 5cm}]{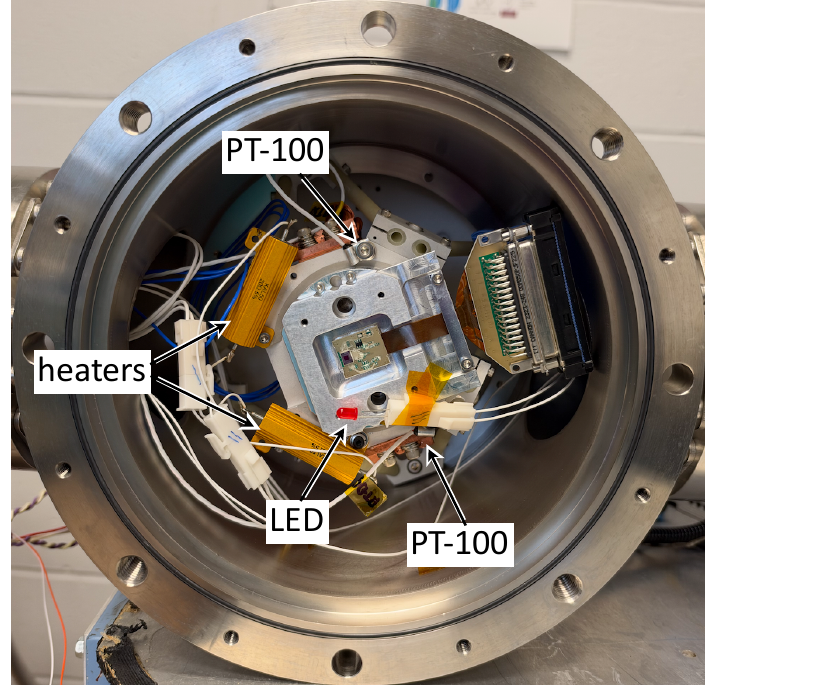}\hfill
\caption{\GeCCD\ prototype in the BNL cryogenic test system. Vacuum chamber with the
packaged device on the cold plate and the CCD controller connected through the short
external cable. The device is mounted in the chamber with the package cover removed.
}
\label{fig:setup}
\end{figure}

The packaged device is mounted on a cold plate inside a vacuum chamber as shown in
Fig.~\ref{fig:setup}. The cold plate is cooled with liquid nitrogen through copper
thermal links, and the aluminum block shown in Fig.~\ref{fig:packaged_geccd}
is attached to the cold plate with a thin indium foil in between.
The rate of temperature change during cooldown and warm-up was limited to
3.5\,K min$^{-1}$ to minimize the mechanical stress on the CCD arising from the
mismatch in thermal expansion coefficients between the device and its package.
The CCD reaches a base temperature of about 95\,K, and its temperature was
stable to within $\pm0.03$\,K during the dark-current measurements.
The chamber pressure was below $5\times10^{-5}$\,torr.  

The CCD clocks and biases are supplied, and the video signal is digitized,
by an Archon CCD controller~\cite{archon_controller}, using test infrastructure
adapted from LSST development systems~\cite{lsst_poc_2008, lsst_poc_2016}.
Table~\ref{table:clock_voltages} lists the clock voltages applied to the parallel
and serial gates and to the reset gate. 
During the exposure, two of the three parallel gates are held at $-2.0$\,V
and the third at 0.0\,V, instead of one gate at $-2.0$\,V and two
at the $+1.0$\,V level used for charge transfer.
Holding two gates low maximizes the depletion depth, and lowering
the barrier gate from $+1.0$ to 0.0\,V reduces the dark current.
The cables between the device and the controller are kept short,
about 100\,mm inside the chamber and 70\,mm outside it.

\begin{table}[htbp]
\centering
\caption{Clock levels.}
\label{table:clock_voltages}
\small
\begin{tabular}{lcc}
\hline
Clock & High (V) & Low (V) \\
\hline
Parallel gates (transfer) & $+1.0$ & $-2.0$ \\
Parallel gates (exposure)$^{a}$ & $\phantom{+}0.0$ & $-2.0$ \\
Serial gates   & $+1.0$ & $-2.0$ \\
Reset gate     & $+2.0$ & $-0.5$ \\
\hline
\end{tabular}

\smallskip
{\footnotesize $^{a}$ Two of the three gates at the low level, one at the high level.}
\end{table}

\subsection{Readout and signal processing}

The device is read out in one of two modes: an unbinned ($1\times1$) mode
and a $2\times2$ on-chip binning mode as shown in Table~\ref{table:readout_mode}.
In the unbinned mode, each frame is read out as $300\times300$ pixels,
so that it contains the active area together with serial and parallel overclock regions,
which are used to determine the baseline, the readout noise, and the dark current.
The parallel clocks run at 2\,kHz and the serial register at about 29\,kHz,
resulting in a frame readout time of 3.25\,s.

In the $2\times2$ mode, the charges of two rows are transferred
into the serial register before each line is read out,
and the charges of two serial pixels are combined at the output node
before each sample, so that the four pixels are read as one
$16.2\times16.2$\,\textmu m$^2$ pixel.
The frame then consists of $150\times150$ binned pixels,
and the frame readout time is reduced to 1.14\,s.
Because the charge is summed before it reaches the output amplifier,
binning does not increase the readout noise per binned pixel,
whereas summing the same four pixels after digitization would
increase it by a factor of two.
The dark charge per binned pixel, however, is the sum of the four pixels.

For the X-ray measurements, the exposure time between readouts was
set to 51\,ms. Because no shutter is used, the device remains sensitive
during readout, and the position of an X-ray interaction along the
parallel-transfer direction cannot be determined.

\begin{table}[htbp]
\centering
\caption{Readout modes.}
\label{table:readout_mode}
\begin{tabular}{lcc}
\hline
 & Unbinned ($1\times1$) & $2\times2$ binning \\
\hline
Effective pixel size (\textmu m) & 8.1 & 16.2 \\
Frame format (pixels) & $300\times300$ & $150\times150$ \\
Pixel period (\textmu s) & 34.1 & 43.7 \\
Frame readout time (s) & 3.25 & 1.14 \\
\hline
\end{tabular}
\end{table}

\begin{figure}[htbp]
\centering
\includegraphics[width=\linewidth, clip, trim={.5cm 6.5cm .5cm 7cm}]{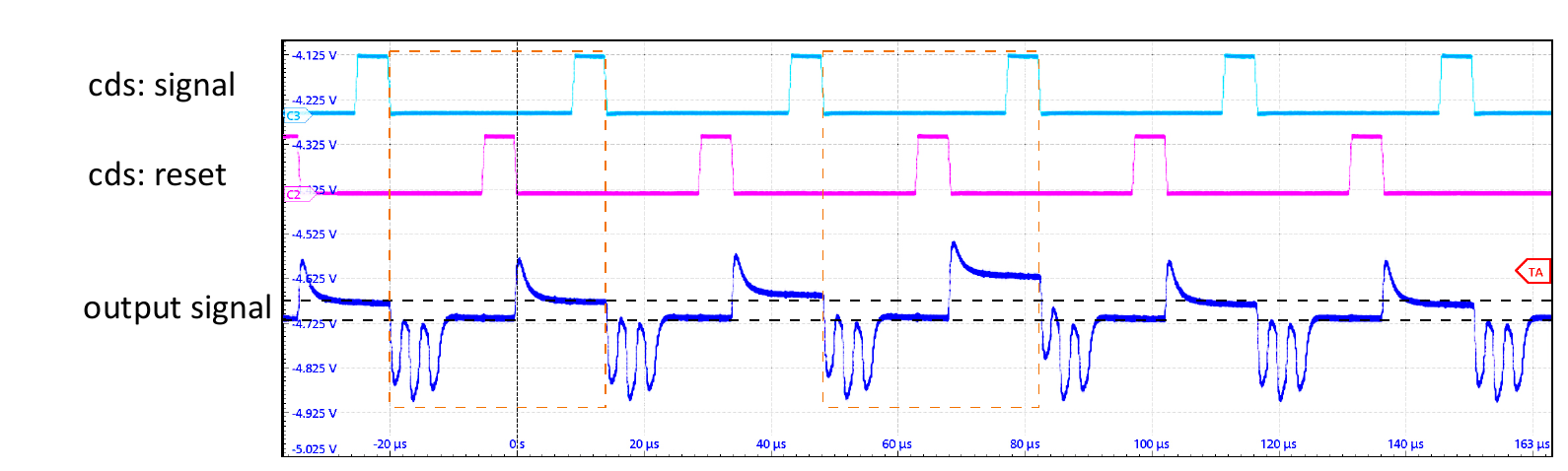}
\caption{Oscilloscope capture of the CCD output (dark blue, 100~mV per division)
    and two clock signals making the CDS integration for reset level (magenta)
    and signal level (cyan) over several pixel cycles. The pixel
    period, shown by broken square, is about 34.1~\textmu s.
    The horizontal dashed lines mark the two plateaus that are
    sampled for correlated double sampling, and the dashed boxes mark individual pixel cycles. The cycle in the second box has a different signal plateau from its neighbors, as expected
for a pixel that contains signal charge for an X-ray hit.}
\label{fig:waveform}
\end{figure}

Figure~\ref{fig:waveform} shows the output waveform over several pixel cycles for the unbinned mode.
Each cycle
has a period of about 34\,\textmu s and contains two plateaus, a reference level and a signal
level, each of which follows a clock edge and its feedthrough transient. The controller
samples both plateaus and takes their difference (correlated double sampling, CDS), which
gives the pulse height of the pixel in analog-to-digital units (ADU). Both plateaus last
for several microseconds after the transients have decayed, which leaves room for the two
sampling windows of 5.12\,\textmu s, and a pixel that contains signal charge is visible directly in the waveform as a shift of its signal plateau.

\section{Measurement Methods}
\label{sec:methods}

\subsection{Dark current}

Dark frames were taken at exposure times between 1 and 20~s,
at a CCD temperature of
$94.69\pm0.04$~K, with a readout time of 3.25~s per frame.
Hot pixels were identified in the 20~s frame as pixels
whose signal exceeded the mean by more than $3\sigma$,
where $\sigma$ is the standard deviation of the pixel value.
The same hot pixels were excluded from the active area
for all exposure times.
For each exposure the dark
signal is the difference between the signal level of the active area and that of the
vertical overclock region, converted to electrons with the conversion gain. The
dark-current rate is the slope of a straight-line fit of dark signal against exposure time,
and its quoted uncertainty is the $1\sigma$ uncertainty of the fitted slope.

\subsection{X-ray measurements and event selection}

The CCD chamber is equipped with an X-ray entrance window made of a 200-nanometer-thick silicon nitride membrane that transmits X-rays down to the C-K line (0.28\,keV).
The device was irradiated with three radioactive sources: \Fe\ (Mn~\Ka\ at 5.90~keV and
Mn~\Kb\ at 6.49~keV), \Ba\ (Cs~\Ka\ at 30.9~keV and Cs~\Kb\ at 35.1~keV),
and \Am\ (59.5~keV).
Each source was placed in front of the X-ray entrance window. The distance between the source and the window was adjusted so that the event pile-up was low enough not to affect the measured spectra.

The charge generated by X-rays below 10 keV is mostly confined to a single pixel
or a $2 \times 2$ pixel region. In contrast, that generated by X-rays above 30\,keV
spreads over $3 \times 3$ pixels or more because these X-rays are absorbed deeper
in the device and produce longer photoelectron tracks.
Therefore, the \Fe\ data were taken in the unbinned mode
and the \Ba\ and \Am\ data were taken in the $2 \times 2$ binning mode
shown in Table~\ref{table:readout_mode}, which collects the charge of
such events in fewer, larger pixels without adding readout noise.

An averaged dark frame was obtained by averaging 100 dark frames and was subtracted from each frame taken under X-ray irradiation. X-ray events were then extracted using two thresholds, an event threshold and a split threshold. A pixel was identified as the center of an X-ray event when its pulse height exceeded the event threshold and was the local maximum within the surrounding $3 \times 3$ pixels. The pulse heights of the eight neighboring pixels were then compared with the split threshold, and charge was considered to have spread into every neighboring pixel whose pulse height exceeded it. Each event was classified according to the resulting charge distribution, using two grading schemes: the ASCA grade and the Chandra ACIS grade~\cite{asca_chandra_grade}. The event threshold was set to 50\,ADU. The split threshold was set to three times the standard deviation of the pulse heights in the horizontal overclock region, which was 5.0 and 5.4\,ADU, resulting in split thresholds of 15 and 16\,ADU, for the unbinned and $2 \times 2$ binning modes, respectively.
The histogram was rebinned so that each bin contained at least 20 counts, allowing
$\chi^2$ statistics to be applied.
For \Fe, only single-pixel (grade~0) events were used; for \Ba,
events of ASCA grades 0, 2, 3, 4, and 6; and for \Am, events of all grades.

\subsection{Conversion gain and readout noise}
\label{sec:gain_method}

The conversion gain is calibrated with X-ray lines of known energy. A straight-line fit of
the fitted peak position against photon energy gives the energy scale $S$ in eV ADU$^{-1}$. The
conversion gain $K$ in \el~ADU$^{-1}$ then follows from the mean energy $w$ needed to create an
electron--hole pair,
\begin{equation}
  K = \frac{S}{w},
\end{equation}
with $w = 2.97$~eV for germanium at 77\,K~\cite{ge_pair_creation}.

The readout noise is determined as the standard deviation of the pixel values in a $5\times235$‑pixel region of the horizontal overclock, which contains no signal charge and negligible dark charge, multiplied by $K$. 

\section{Results}
\label{sec:results}

\subsection{Dark current}

\begin{figure}[htbp]
\centering
\includegraphics[width=0.62\linewidth, clip, trim={0 .5cm 0 10cm}]{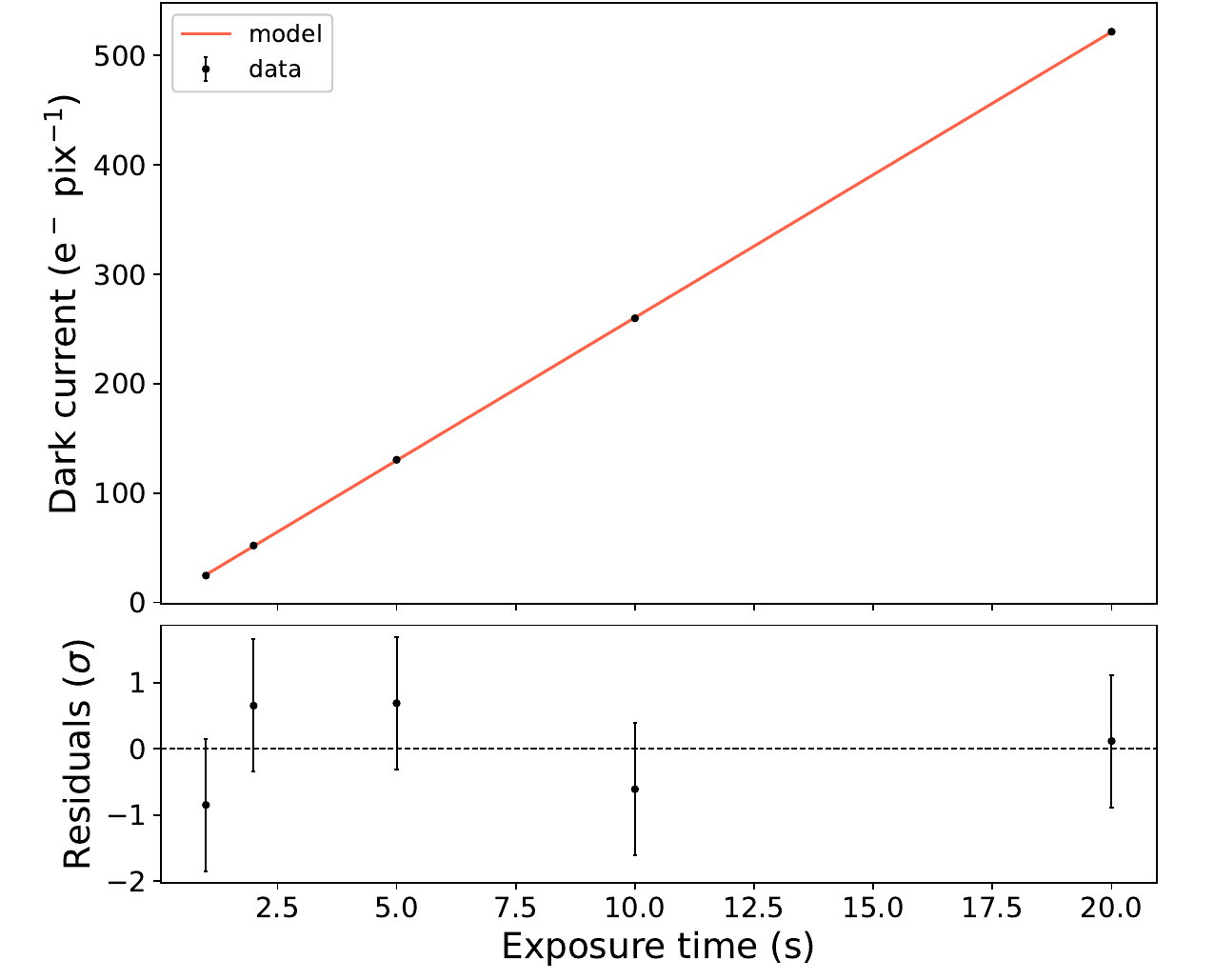}
\caption{Dark signal of device \dut\ as a function of exposure time at    
    $94.69 \pm 0.04$~K.
    The red solid line shows the best-fit linear function. The lower panel shows
    the fit residuals normalized by the statistical uncertainties
    of the data errors. The fitted slope is
    $26.12 \pm 0.08$~\elpps.}
\label{fig:dark_current}
\end{figure}

Figure~\ref{fig:dark_current} shows the dark signal as a function of exposure time. The
signal increases linearly with exposure time, and the fitted slope gives a dark current of
$26.12 \pm 0.08$~\elpps\ at $94.69 \pm 0.04$~K.

\subsection{X-ray response}\label{sec:xray_response}

\begin{figure}[htbp]
\centering
\includegraphics[width=0.48\linewidth]{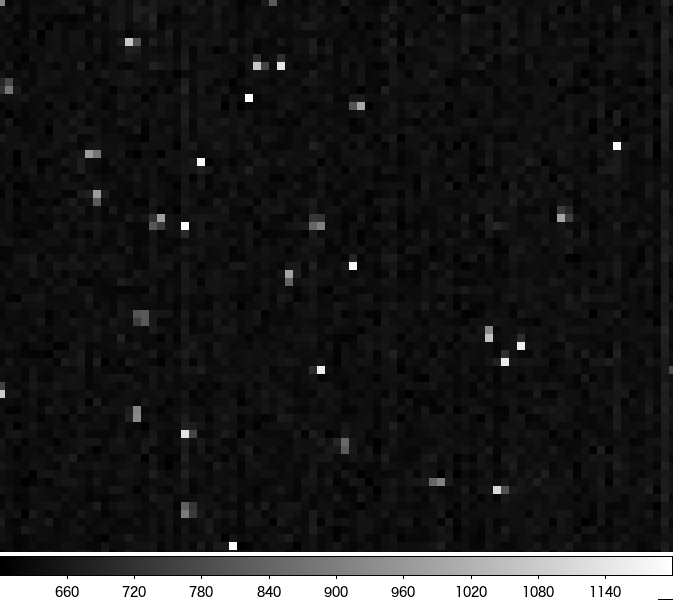}\hfill
\includegraphics[width=0.48\linewidth, clip, trim={1cm .5cm 0 12cm}]{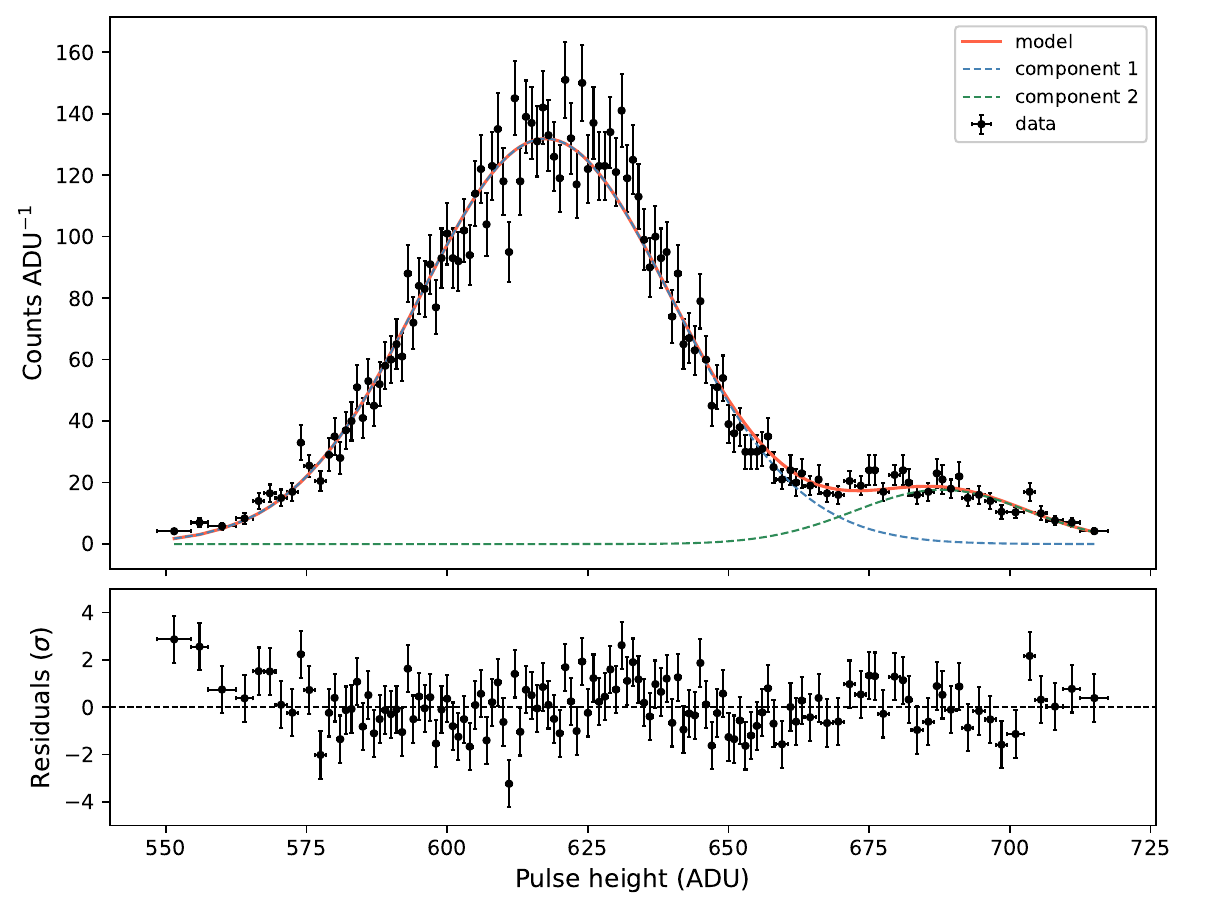}
\includegraphics[width=0.5\linewidth, clip, trim={7cm 4.3cm 5cm 5cm}]{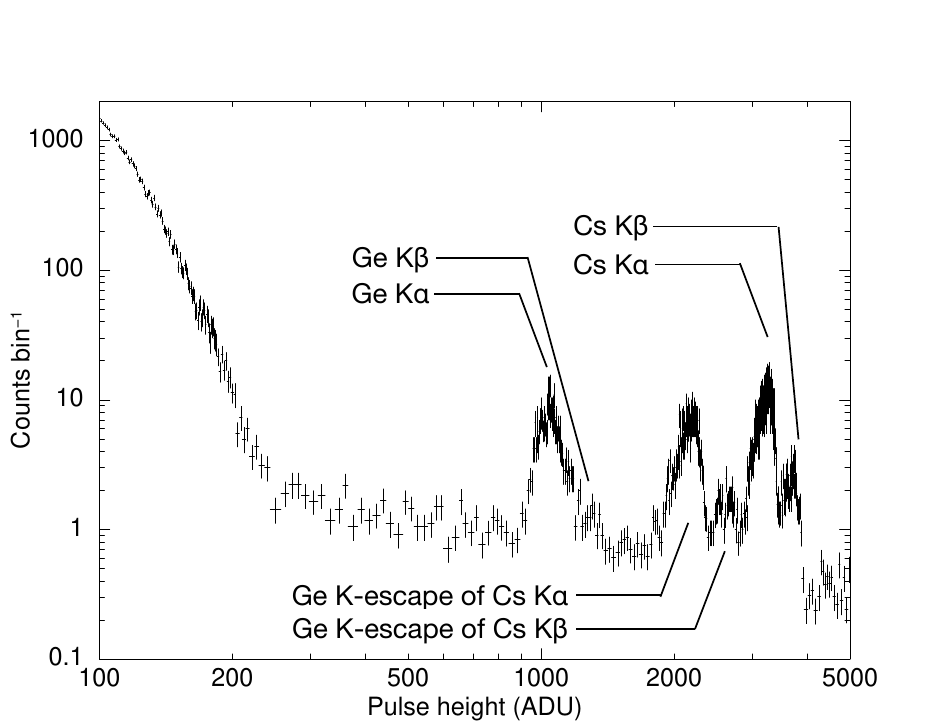}\hfill
\includegraphics[width=0.5\linewidth, clip, trim={7cm 4.3cm 5cm 5cm}]{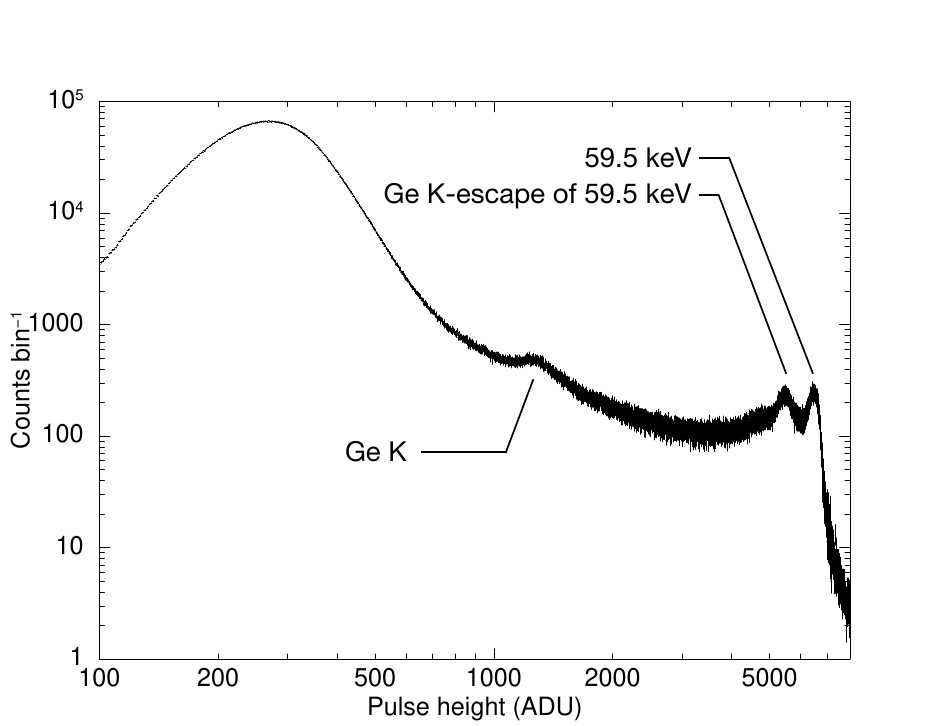}
\caption{X-ray response of device \dut.
    Top left: raw frame under \Fe\ irradiation at
    94.7~K. Top right: Pulse-height spectrum of single-pixel events
    obtained under $^{55}$Fe
    irradiation. The red solid line shows the best-fit model,
    the sum of two Gaussians, and the dashed lines show the individual
    Mn K$\alpha$ and Mn K$\beta$ components.
    The lower panel shows the fit residuals normalized by the data error.
    Bottom: ({\it left}) \Ba\ and ({\it right}) \Am\ pulse-height spectra.
    Each line feature is identified on the plot.}
\label{fig:xray}
\end{figure}

Figure~\ref{fig:xray} (top left) shows a raw CCD image obtained under \Fe\ irradiation.
The charge generated by a Mn K X-ray is mostly confined to a single pixel and spreads
over at most $2 \times 2$ pixels. The pulse-height histogram of the single-pixel (grade~0)
events is shown in Fig.~\ref{fig:xray} (top right), in which the Mn \Ka\ and Mn \Kb\
lines are resolved. After rebinning, the histogram was fitted with two Gaussian
functions. The energy resolution, expressed as the full width at half maximum (FWHM),
is $8.5\pm0.1$\,\% for Mn \Ka\ and $5.3\pm0.3$\,\% for Mn \Kb.

Figure~\ref{fig:xray} shows the pulse-height histograms obtained
under \Ba\ (bottom left) and \Am\ (bottom right) irradiation.
The CCD was operated with $2\times2$ on-chip binning, and
the histograms were made from events of ASCA grades 0, 2, 3, 4, and 6
for \Ba\ and from events of all grades for \Am.
The identified lines are labeled in each panel.
The \Ba\ histogram shows the Cs \Ka\ and \Kb\ lines,
together with the Ge K-escape peaks of the Cs lines.
The \Am\ histogram shows the 59.5\,keV line and its Ge K
escape peak. Ge K fluorescence lines, produced by Ge K X-rays that escape from the
primary interaction site and are absorbed elsewhere in the device, are present in
both histograms.
Note that the Ge K fluorescence lines in the \Am\ histogram are broadened
because events of all grades are accumulated.
In the histograms obtained with \Ba\ and \Am\ irradiation,
the continuum below the Ge K line arises from noise, scattered X-rays, and X-rays
absorbed in the field-free or low-field region below the depletion layer.
In the latter case, the charge cloud diffuses beyond the $3\times3$ pixel window,
so only part of the charge is collected.
The FWHM is $8.4\pm0.4$\,\% for the Cs \Ka\ line at 30.9\,keV and
$7.45\pm0.04$\,\% for the 59.5\,keV line. Table~\ref{tab:results} summarizes these
measurements.

\begin{table}[htbp]
\centering
\caption{Summary of the measurements of device \dut.}
\label{tab:results}
\small
\begin{tabular}{lll}
\hline\hline
Quantity & Value & Conditions \\
\hline
Dark current & $26.12 \pm 0.08$~\elpps & $94.69 \pm 0.04$~K \\
Energy scale & 9.15~eV ADU$^{-1}$ & X-ray line calibration \\
Conversion gain & 3.08~\el ADU$^{-1}$ & $w$\,=\,2.97~eV \\
Readout noise & 15.4~\el\ rms & 5.02~ADU in overclock with unbinned mode \\
FWHM at 5.9~keV (\Fe) & $8.5 \pm 0.1$\% & Single-pixel events \\
FWHM at 30.9~keV (\Ba) & $8.4\pm0.4$\% & Grade 02346 \\
FWHM at 59.5~keV (\Am) & $7.45\pm0.04$\% & All grades \\
\hline
\end{tabular}
\end{table}

\subsection{Conversion gain and readout noise}

\begin{figure}[htbp]
\centering
\includegraphics[width=0.5\linewidth, clip, trim={0 .5cm 0 10cm}]{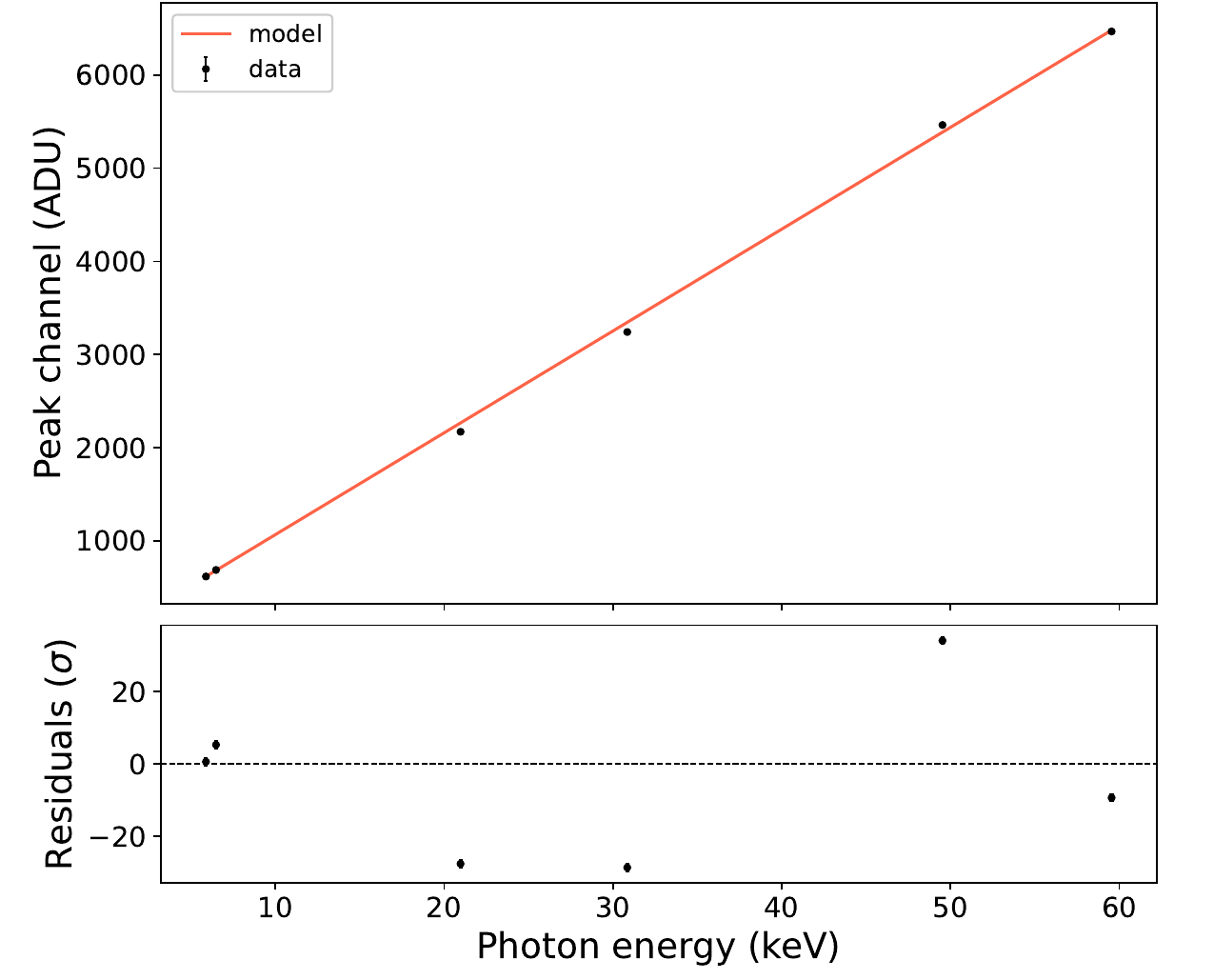}
\caption{Peak channel as a function of photon energy. The red solid line
    shows the best-fit linear function. The lower panel shows
    the fit residuals normalized by the statistical uncertainties
    of the peak positions.}
\label{fig:e-ph}
\end{figure}

 The peak channels of the lines in the X-ray spectra obtained with $^{55}$Fe,
$^{133}$Ba and $^{241}$Am were determined by fitting Gaussian functions.
Figure~\ref{fig:e-ph} shows the peak channel as a function of photon energy.
A linear fit to these data yields a slope of 109.5\,ADU keV$^{-1}$, corresponding to
an energy scale of 9.15\,eV ADU$^{-1}$.
With a pair-creation energy of $w$\,=\,2.97\,eV~\cite{ge_pair_creation},
the conversion gain is 3.08\,e$^-$ ADU$^{-1}$.
The standard deviation of the horizontal overclock is
5.02~ADU, so the readout noise is 15.4~\el\ root mean square (rms).

\section{Discussion}
\label{sec:discussion}

\subsection{Interpretation of the prototype measurements}\label{sec:interpretation}

The measurements show that the BNL test system can operate the packaged \GeCCD\ near 95~K,
read it out with correlated double sampling, and calibrate its response in absolute units.
The measured dark current of 26.12\,\el\,pixel$^{-1}$\,s$^{-1}$ or 6.4\,pA cm$^{-2}$ at 94.69\,K is far higher than expected from bulk generation. The estimate from \cite{leitz2017} for a fully depleted 50\,\textmu m-thick Ge-CCD extrapolates to about $10^{-7}$\,pA cm$^{-2}$ at 95\,K and reaches the measured value only near 150\,K. 
Ref.~\cite{leitz2020} discusses two possible causes in Ge-CCDs. The first is parasitic MOSFETs, which flatten the temperature dependence of the dark current below approximately 150\,K. The second is bulk traps, revealed by first-pixel-edge-response (FPER) measurements of charge-transfer inefficiency. The same or similar traps that degrade charge transfer can also act as generation centers and increase the dark current. Further diagnosis, including measurements of the dark current as a function of temperature, is needed to determine which mechanism dominates in this device.

The confinement of \Fe\ events to $2\times2$ pixels shows that signal charge is collected
and transferred as compact packets. 
 The energy resolution is not yet limited by readout noise or by
 charge-generation statistics.
 At 5.9~keV, where only single-pixel events are used, the readout noise of
 15.4~\el\ rms contributes about 0.11~keV FWHM,
 and the statistical (Fano) limit is about 0.10~keV for a Fano factor of 0.112~\cite{ge_fano_factor}.
 The two together account for
 about 0.15~keV, or 2.5\%, against the measured 8.5\%.
 For \Ba, events of ASCA grades of 0, 2, 3, 4, and 6 are used and the pulse height is
 mostly the sum of up to $2 \times 2$ pixels for 30.9 keV,
 so the relevant readout noise is as large as $16.6\times\sqrt{4}=33.3$\,\el\ rms,
 or 0.23\,keV FWHM.
 Similarly, for \Am, events of all grades are used and the pulse height is
 mostly the sum of up to $3 \times 3$ pixels for 59.5 keV,
 so the relevant readout noise is as large as $16.6\times\sqrt{9}=49.9$~\el\ rms,
 or 0.35\,keV FWHM.
 Even with this upper limit, readout noise and Fano statistics
 account for only 0.33~keV (1.1\%) at 30.9~keV and 0.48~keV (0.8\%)
 at 59.5~keV, against the measured 8.4\% and 7.45\%.

 Two contributions are missing from this estimate.
 The first is dark current. Because the exposure time (51~ms)
 is much shorter than the frame readout time (3.25\,s for no binning
 and 1.14\,s for on-chip $2 \times 2$ binning mode), nearly all of
 the dark charge in a packet is accumulated
 while the charge packet is transferred through the other pixels of its column.
 The averaged dark frame therefore represents the dark signal
 averaged along the column, and subtracting it does not correct
 the dark signal pixel by pixel.
 In particular, the shot noise of the dark charge accumulated
 during the $\approx$2.8\,s for \Fe\ and $\approx$0.97\,s for \Ba\ and \Am\ 
 that a packet spends in the imaging area
 remains in every pixel. This noise is absent from the overclock region,
 where the readout noise was measured, and adds to the 15.4~\el.
 The second is a broadening that scales with the deposited energy,
 which points to a variation in the fraction of charge that is
 collected and transferred, for example from incomplete charge collection,
 charge-transfer inefficiency (CTI), or
 charge lost below the split threshold.
 As the device remains sensitive during readout,
 it is impossible to determine the row in which each X-ray
 was absorbed, meaning that parallel CTI cannot be corrected.
 Separating these contributions is a priority for the next measurements.

\subsection{Context within the SWIR detector trade space}

\begin{figure}[htbp]
\centering
\includegraphics[width=\linewidth]{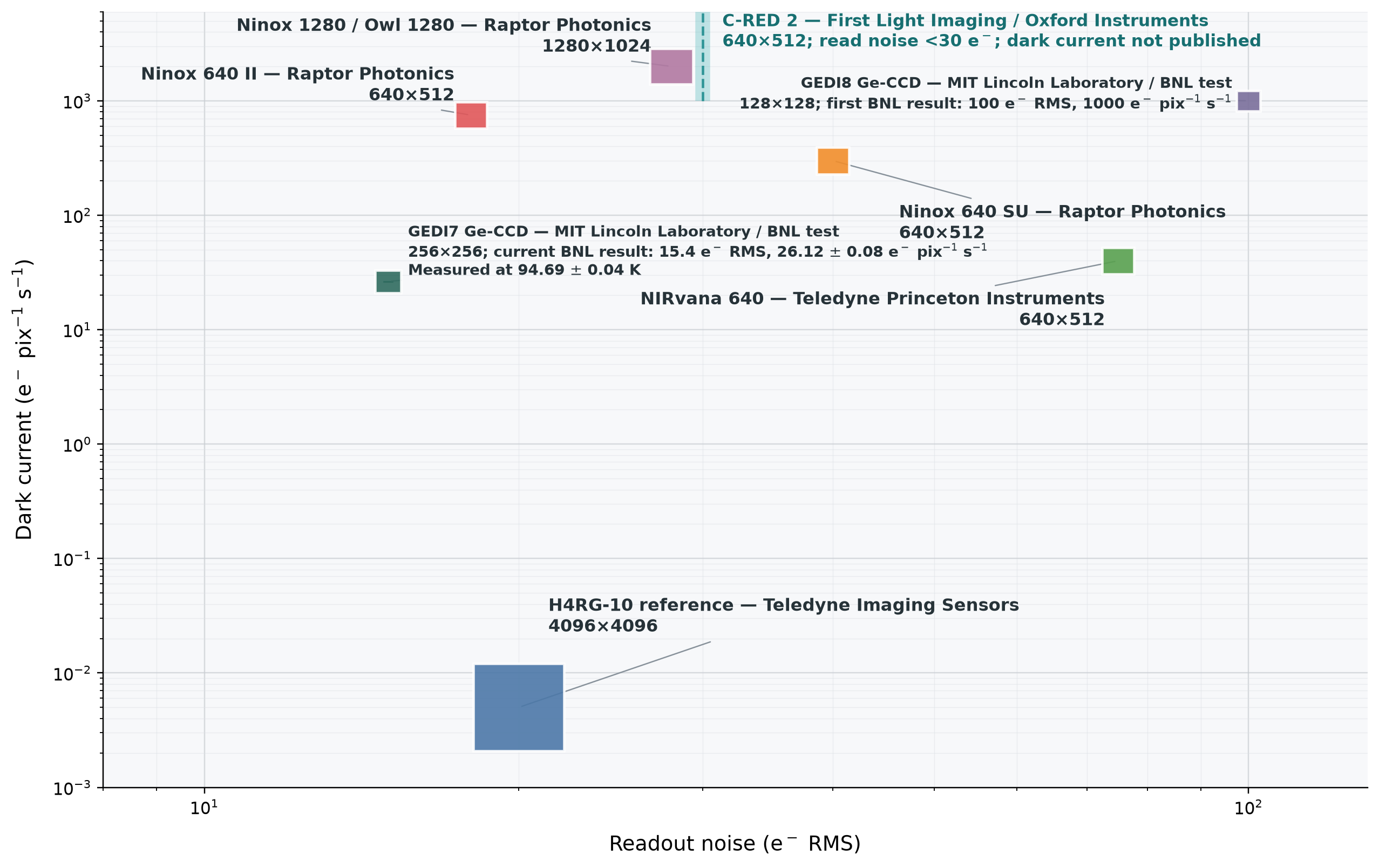}
\caption{Comparison of readout noise and dark current for selected SWIR detector technologies relevant to a spectrograph channel operating near 1.4~\textmu m. Manufacturer specifications are used for the Teledyne H4RG-10 and NIRvana~640, the Raptor Photonics Ninox cameras, and the First Light Imaging C-RED~2~\cite{teledyne_h4rg10_2021,teledyne_nirvana640,raptor_ninox640su,raptor_ninox640ii,raptor_ninox1280,firstlight_cred2}. The two Ge-CCD points show the progression of the BNL test campaign from GEDI8 ($100~\mathrm{e}^-$ rms and $1000~{\mathrm e}^-\,\mathrm{pix}^{-1}\,\mathrm{s}^{-1}$) to GEDI7 ($15.4~{\mathrm e}^-$ rms and $26.12~\mathrm{e}^-\,\mathrm{pix}^{-1}\,\mathrm{s}^{-1}$). The Ge-CCD array format is described in Ref.~\cite{donlon2025geccd}; the performance values shown here are from the BNL measurements reported in this work. Rectangle dimensions represent the relative pixel format of each array. For C-RED~2, the dashed vertical segment indicates that a read-noise specification is available but a directly comparable dark-current value was not used. The comparison is indicative because the sensors differ in pixel size, array format, operating temperature, readout mode, and measurement methodology. 
}
\label{fig:swir_trade_space}
\end{figure}

Figure~\ref{fig:swir_trade_space} places the \GeCCD\ results in the plane of readout noise and
dark current alongside selected SWIR sensors, for a spectrograph channel near
1.4~\textmu m. The device characterized here has a readout noise of
15.4~e$^{-}$ rms and a dark current of
26.12\,e$^{-}$ pixel$^{-1}$ s$^{-1}$ at 94.69\,K, an improvement on
both axes over the first device tested at BNL (GEDI8). The device and the
readout configuration both changed between the two measurements.

These values lie in the range of commercial InGaAs cameras, with two
qualifications. Those cameras operate at much higher temperatures, and their
pixels are several times larger in area than the 8.1~\textmu m pixels of the
\GeCCD, so its dark current per unit area is higher. Astronomy-grade HgCdTe
arrays, represented in the figure by the H4RG-10, have dark currents more
than four orders of magnitude lower. For the long exposures of a
spectrograph, dark current rather than readout noise is therefore the main
gap to close (Section~\ref{sec:interpretation}). Whether it decreases
at lower temperature depends on its origin (Section~5.1),
and measuring it below 95,K is the next step in the characterization
of this device. The comparison is otherwise indicative: the sensors differ in array
format and readout mode, and their published values were not all measured in
the same way.

The interest of the \GeCCD\ lies in its architecture rather than in its
present performance. It is a monolithic device: it needs no infrared
absorber layer hybridized to a CMOS readout circuit, and it moves signal in
the charge domain to a small number of output amplifiers. Charge-domain
operation allows on-chip binning and time-delay integration without added
readout noise; the latter is the requirement of scanning astrometry concepts
such as GaiaNIR~\cite{hobbs2019scanningnir}. Monolithic fabrication may also
lower the cost per pixel, which the studies of an infrared camera for Rubin
Observatory after LSST identify as the condition for a germanium focal plane
to be viable~\cite{stubbs2019lsstnextgen,kahn2019lsstfuture}. These benefits
remain prospective. A lower dark current is the first requirement. The optical response,
the charge-transfer efficiency, the full-well capacity, large-format
scalability, and stability must also be demonstrated.


\subsection{Development priorities toward future near-infrared instrumentation}

The present data are an electrical and X-ray baseline for a front-illuminated
prototype.
The next step is to measure the dark current as a function of temperature below 95\,K, to determine whether it is limited by parasitic MOSFETs or by bulk traps (Section.~\ref{sec:interpretation}).
The charge-transfer
efficiency, the optical response and quantum efficiency across the intended
band, and the stability of the device under repeated operation follow.

Two developments are needed to turn the prototype into a detector for an
instrument. The first is backside processing. In a front-illuminated device
the light crosses the gate structure before it reaches the germanium, which
limits the quantum efficiency. A science-grade device must be illuminated
from the back, which requires a thin, stable backside contact and an
anti-reflection coating matched to the near-infrared band. Backside
processing of germanium is therefore a central part of the program.

The second is operation with full depletion. A backside contact allows a
substrate voltage to be applied across the device, so that the whole
thickness is depleted and charge generated near the back surface drifts to
the buried channel instead of diffusing. Full depletion controls the lateral
spread of the charge, which sets the spatial resolution of an imager and the
spectral resolution of a spectrograph, and it is the approach that gave thick
silicon CCDs their response in the red~\cite{holland2003}. The measurements
reported here were made below full depletion. Operating the present device at
higher substrate voltage is a near-term test: it will show how the dark
current and the charge collection depend on the depleted volume, and whether
the broadening of the X-ray lines discussed in
Section~\ref{sec:xray_response} comes from charge generated in undepleted
material.

The main application we have in view is Spec-S5. Its baseline spectrographs
cover 360--980~nm in three channels with silicon CCDs, and an additional
near-infrared channel would extend the coverage to about 1.4~\textmu m and give
access to galaxies in the redshift range $1.5 \le z \le
2.0$~\cite{besuner2025}. Such a channel needs low readout noise and low dark
current over long exposures, high quantum efficiency out to its red limit,
and formats comparable to those of the silicon channels. The results of this
paper and the measurements listed above are steps toward a comparison with
those requirements. Dedicated low-noise cryogenic readout electronics are
part of the same effort, and multiple-sampling output stages in germanium are
a longer-term direction~\cite{stubbs2019lsstnextgen}.

Laboratory characterization will not be sufficient on its own. Before a
germanium detector is adopted for a survey instrument, a packaged,
back-illuminated device should be operated on the sky in a pathfinder
instrument, where sky subtraction, calibration, and long-term operation can
be tested under observing conditions. Pathfinder projects of this kind have
been identified as the route by which germanium devices could enter Spec-S5,
including as a later spectrograph upgrade~\cite{snowmassCF4}. BNL, MIT
Lincoln Laboratory, and LBNL together cover device fabrication, processing,
readout, and characterization, so that the results of each test campaign can
feed back into the next device iteration.



\section{Conclusions}

We have operated a germanium CCD fabricated at MIT Lincoln Laboratory in a
cryogenic test system at BNL and measured its dark current, readout noise,
and X-ray response. The device, a front-illuminated $256\times256$ p-channel
prototype, is read out with correlated double sampling at a readout noise of
15.4~\el\ rms, with the charge scale calibrated by X-ray lines between 5.9
and 59.5~keV. X-ray events from \Fe\ are confined to at most $2\times2$
pixels, which shows that signal charge is collected and transferred as
compact packets.

Two results define the work ahead. The first is the dark current,
$26.12 \pm 0.08$~\elpps\ at 94.69~K. Its shot noise exceeds the readout
noise after about 0.58 second of exposure, so dark current, not readout noise,
is what separates this prototype from a detector for long astronomical
exposures.
It is far above the bulk-generation estimate, possibly because of
parasitic MOSFETs or bulk traps, and measuring its temperature dependence
below 95\,K is the next step toward identifying its origin.
The second is the energy resolution, 8.5\% FWHM
at 5.9~keV for single-pixel events, 8.4\% at 30.9\,keV
for ASCA grades 0, 2, 3, 4, and 6,
and 7.45\% at 59.5~keV for
all-grade events. It
is well above
the 2.5\% expected at 5.9~keV from readout noise and charge-generation
statistics, which points to incomplete charge collection or transfer.
Operating the device at higher substrate voltage and measuring the
charge-transfer efficiency will test this.

These measurements give a baseline for the device and a working test system
for the next ones. The interest of the \GeCCD\ for near-infrared astronomical
instruments lies in its architecture: a monolithic detector that moves signal
in the charge domain, suited to a near-infrared spectrograph channel for
surveys such as Spec-S5 and to time-delay integration for scanning
astrometry. Reaching that goal requires a lower dark current, backside
processing and fully depleted operation, a measured optical response in the
near infrared, and finally operation on the sky in a pathfinder instrument.
The results reported here are the starting point for that program.

\section*{Acknowledgments}

This work was supported by the Brookhaven National Laboratory
Program Development fund (PD26-006). 
We gratefully acknowledge support for this device fabrication from NASA’s Astrophysics Division through Astrophysics Research and Analysis Award 80HQTR19T0048.
We acknowledge the technical support of A. Verderosa, J. Pinz, J. Procida, R. Lin,
T. Kersten, and D. Pinelli from Brookhaven National Laboratory.

DISTRIBUTION STATEMENT A. Approved for public release. Distribution is unlimited. This material is based upon work supported by the National Aeronautics and Space Administration under Air Force Contract No. FA8702-15-D-0001. Any opinions, findings, conclusions or recommendations expressed in this material are those of the author(s) and do not necessarily reflect the views of the National Aeronautics and Space Administration

\bibliographystyle{unsrt}
\bibliography{references}

\end{document}